\documentclass{ptephy_v1}

\usepackage{amsmath,amssymb}
\usepackage{graphicx}
\usepackage{hyperref}
\makeatletter
\def\caption@documentclass{standard}
\makeatother
\usepackage[compatibility=false]{caption}
\usepackage{subcaption}
\usepackage{multirow}
\usepackage{xcolor}
\usepackage[normalem]{ulem}

\makeatletter
\renewcommand\p@subfigure{\thefigure}
\makeatother

\makeatletter
\def\titlepageheadline{\hbox to \textwidth{\hfil}}
\def\titlepagefootline{\hbox to \textwidth{\hfil}}
\makeatother
\begin{document}
\raggedbottom
\pagestyle{myheadings}

\title{Nonproportional Response of a GAGG Scintillator to Beta and Gamma Radiation over a Wide Energy Range in PIKACHU}
\shorttitle{Nonproportional response of GAGG Scintillator}


\author[1,*]{Takumi Omori}
\shortauthorlist{Omori et al.}
\affil[1]{Graduate School of Science and Technology, University of Tsukuba,  Tsukuba, Ibaraki, 305-8571, Japan 
\email{omori@hep.px.tsukuba.ac.jp}}

\author[2]{Takashi Iida}
\affil[2]{Institute of Pure and Applied Sciences, University of Tsukuba,  Tsukuba, Ibaraki, 305-8571, Japan}

\author[3]{Yosuke Suga}
\affil[3]{Mirion Technologies (Canberra) KK, Taito, Tokyo, 111-0053, Japan}

\author[4,5]{Masao Yoshino}
\affil[4]{Institute for Materials Research, Tohoku University, Sendai, Miyagi, 980-8577, Japan}
\affil[5]{C\&A Corporation, Sendai, Miyagi, 980-0811, Japan}

\author[6]{Ken-Ichi Fushimi}
\affil[6]{Division of Science and Technology, Tokushima University, Tokushima, Tokushima, 770-8506, Japan}

\author[7,8]{Azusa Gando}
\affil[7]{Department of Human Science, Obihiro University of Agriculture and Veterinary Medicine, Obihiro, Hokkaido, 080-8555, Japan}
\affil[8]{Research Center for Neutrino Science, Tohoku University, Sendai, Miyagi, 980-8578, Japan}

\author[9,2]{Nobuo Hinohara}
\affil[9]{Center for Computational Sciences, University of Tsukuba, Tsukuba, Ibaraki, 305-8577, Japan}

\author[8]{Keishi Hosokawa}

\author[1]{Motonao Ishigami}

\author[10, 5]{Kei Kamada}
\affil[10]{New Industry Creation Hatchery Center, Tohoku University, Sendai, Miyagi, 980-8579, Japan}

\author[11]{Tadafumi Kishimoto}
\affil[11]{Research Center for Nuclear Physics, The University of Osaka, Ibaraki, Osaka, 567-0047, Japan}

\author[8]{Keita Mizukoshi}


\author[1]{Takuya Sako}

\author[5]{Yasuhiro Shoji}

\author[1]{Shota Sugi}

\begin{abstract}
Nonproportional response of scintillation light yield for sub-MeV radiations is a well-known characteristic of inorganic scintillators. In the PIKACHU experiment, a precise understanding of the nonproportional response of the GAGG ($\mathrm{Ce:Gd_3Al_2Ga_3O_{12}}$) scintillator is essential for background modeling. In this study, the nonproportional response of the GAGG scintillator was evaluated for both $\beta$ and $\gamma$ rays. The response for $\beta$ rays was measured using the Compton coincidence technique with a germanium detector, while that for $\gamma$ rays was evaluated using several monoenergetic $\gamma$-ray sources. The measurement results showed different responses for $\beta$ and $\gamma$ rays, with a stronger nonproportional response for $\gamma$ rays than for $\beta$ rays over the energy range of 50--2614~keV. The background model incorporating the measured nonproportional responses reproduced the measured background spectra more accurately compared with the model without the nonproportional response correction. These results demonstrate the importance of precise evaluation of the nonproportional response of the GAGG scintillator which is essential 
not only for background modeling in the PIKACHU experiment but also for accurate energy calibration in scintillator-based experiments.
\end{abstract}

\maketitle


\section{Introduction}

The PIKACHU (Pure Inorganic scintillator experiment in KAmioka for CHallenging Underground sciences) experiment aims to observe the double beta decay of $^{160}$Gd using Ce-doped GAGG ($\mathrm{Ce:Gd_3Al_2Ga_3O_{12}}$) scintillators. GAGG scintillators offer several advantages over the Ce-doped GSO ($\mathrm{Ce:Gd_2SiO_{5}}$) scintillator used in the double beta decay search reported in Ref.~\cite{Danevich2001}. First, large-volume crystals can be grown~\cite{Kochurikhin2020, Kamada2016}, enabling high detection efficiency for searching for rare decays. Second, $\alpha$-ray backgrounds can be rejected using pulse shape discrimination. Third, GAGG scintillators exhibit a high light yield of approximately 60,000 photons/MeV, resulting in excellent energy resolution.

Another key factor in improving the sensitivity is the development of GAGG scintillators with minimal contamination from uranium and thorium decay chains. In 2024, a high-purity GAGG scintillator was developed from the purified raw materials, and radioactive contamination was successfully reduced by approximately one order of magnitude compared with conventional GAGG scintillators~\cite{Omori2024}.

To quantitatively understand the background components in GAGG scintillators from the measured energy spectra, it is necessary to construct a precise background model. Such modeling requires a detailed understanding of the characteristics of GAGG scintillators. In particular, GAGG scintillators are known to exhibit a nonproportional response in scintillation light yield for low-energy radiations~\cite{Sibczynski2018, Kaewkhao2016}, which affects the shape of the model spectra. The current background model of the PIKACHU experiment assumes that the light yield for $\beta$ and $\gamma$ rays below 80~keV is only 70$\%$ of that expected from the proportional response determined in the energy range of 356--2614~keV~\cite{Omori2026}.

In the present study, the nonproportional response of the GAGG scintillator used in the PIKACHU experiment was evaluated for both $\beta$ and $\gamma$ rays. In addition, the measured nonproportional responses were incorporated into the background model to improve its precision.

\section{Experiment} 
In this study, the nonproportional response for $\gamma$ rays was evaluated from photopeak measurements using nine $\gamma$-ray energies emitted by several radioactive sources, whereas that for $\beta$ rays was evaluated using the Compton coincidence technique (CCT) with 511 and 1274~keV $\gamma$ rays from $^{22}$Na and 2614~keV $\gamma$ rays from $^{208}$Tl.

\subsection{Compton coincidence technique}

The CCT is an experimental method used to isolate electrons produced by Compton scattering with well-defined energies. In this technique, monoenergetic $\gamma$ rays are irradiated a test detector, where Compton scattering occurs. The scattered $\gamma$ rays are subsequently detected by a reference detector placed at a fixed angle relative to the incident $\gamma$ rays. By requiring a time coincidence between the signals from these detectors, events corresponding to a specific Compton scattering angle can be selected. The energy of the scattered $\gamma$ ray is determined by the scattering angle according to the Compton formula,

\begin{equation}
E_{\gamma}^{\prime} = \frac{E_{\gamma}}{1 + \frac{E_{\gamma}}{m_e c^2}(1-\cos\theta)},
\label{eq:eq1}
\end{equation}

\noindent
where $E_{\gamma}$ and $E_{\gamma}^{\prime}$ denote the incident and scattered $\gamma$-ray energies, respectively, $m_e c^2$ is the electron rest energy, and $\theta$ is the scattering angle. The CCT enables precise measurements of detector response for low-energy radiations.

\subsection{Experimental setup}
\label{sec:Experimental setup}

\begin{figure}[htbp]
  \centering
  \subcaptionbox{Schematic diagram of the detector arrangement and Nuclear Instrument Module (NIM) electronics setup. \label{fig:Experiment_setup}}{%
    \includegraphics[width=0.75\linewidth]{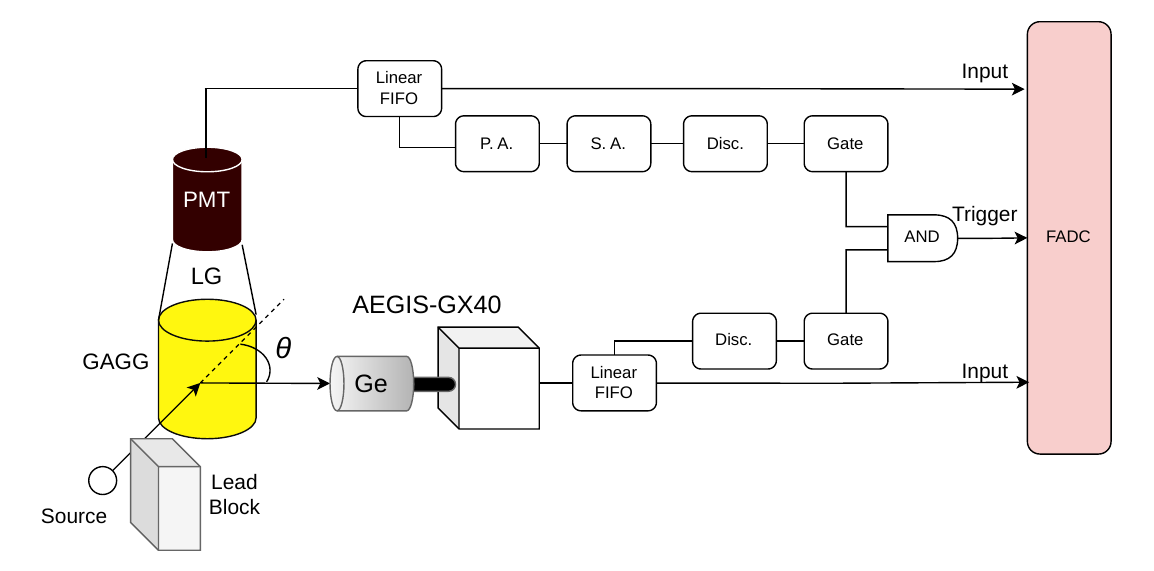}}

  \vspace{0.8em}

  \subcaptionbox{Photograph of the experimental apparatus. The $\gamma$-ray source was placed behind the lead blocks indicated by the white dashed circle. The GAGG scintillator, light guide, and photomultiplier tube were covered with a black sheet for light shielding during the measurements. \label{fig:Experiment_photo}}{%
    \includegraphics[width=0.75\linewidth]{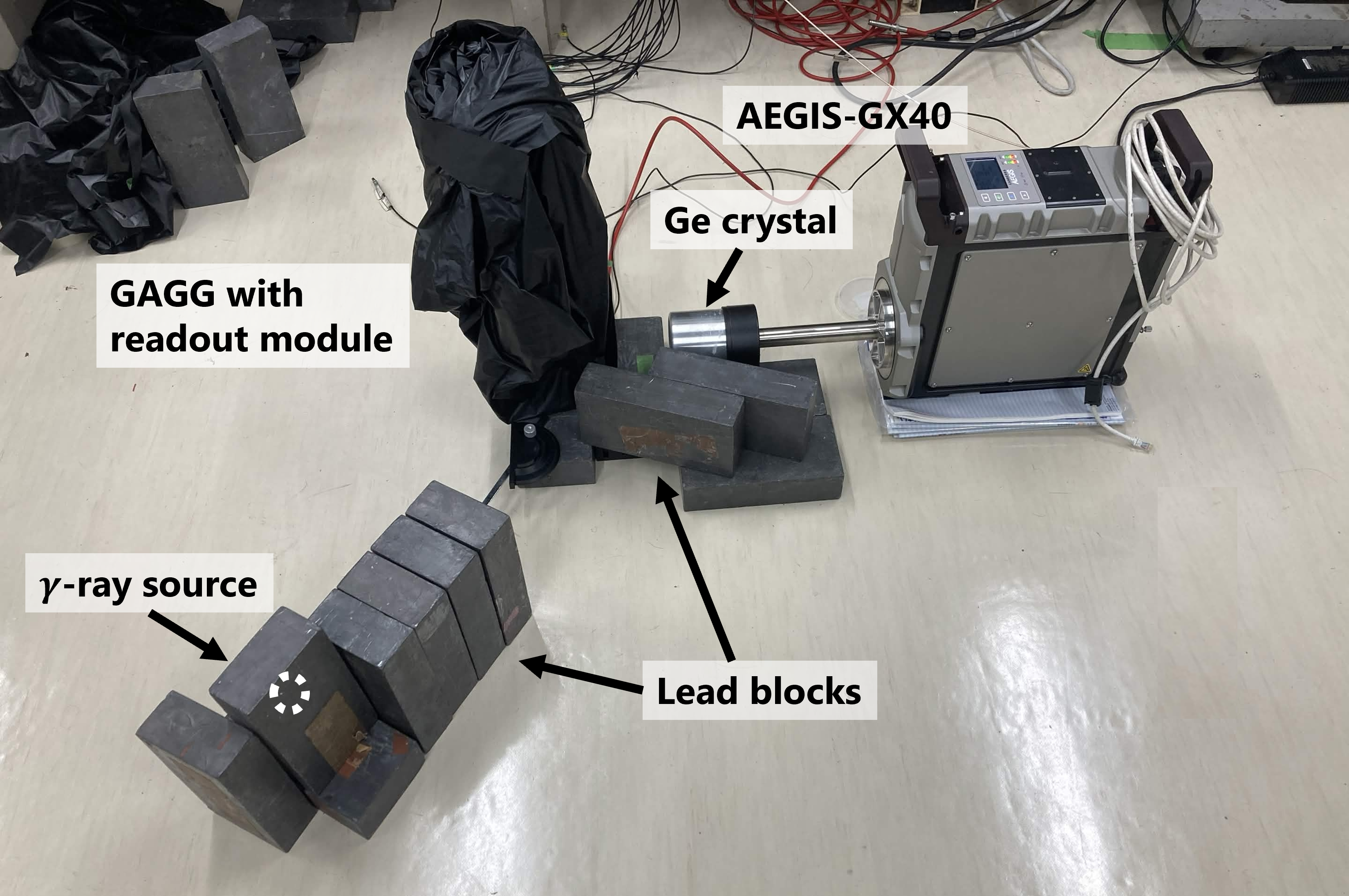}}
  \caption{Experimental setup used in this study.}
  \label{fig:Experiment}
\end{figure}

\begin{figure}[htbp]
  \centering
  \includegraphics[width=0.4\linewidth]{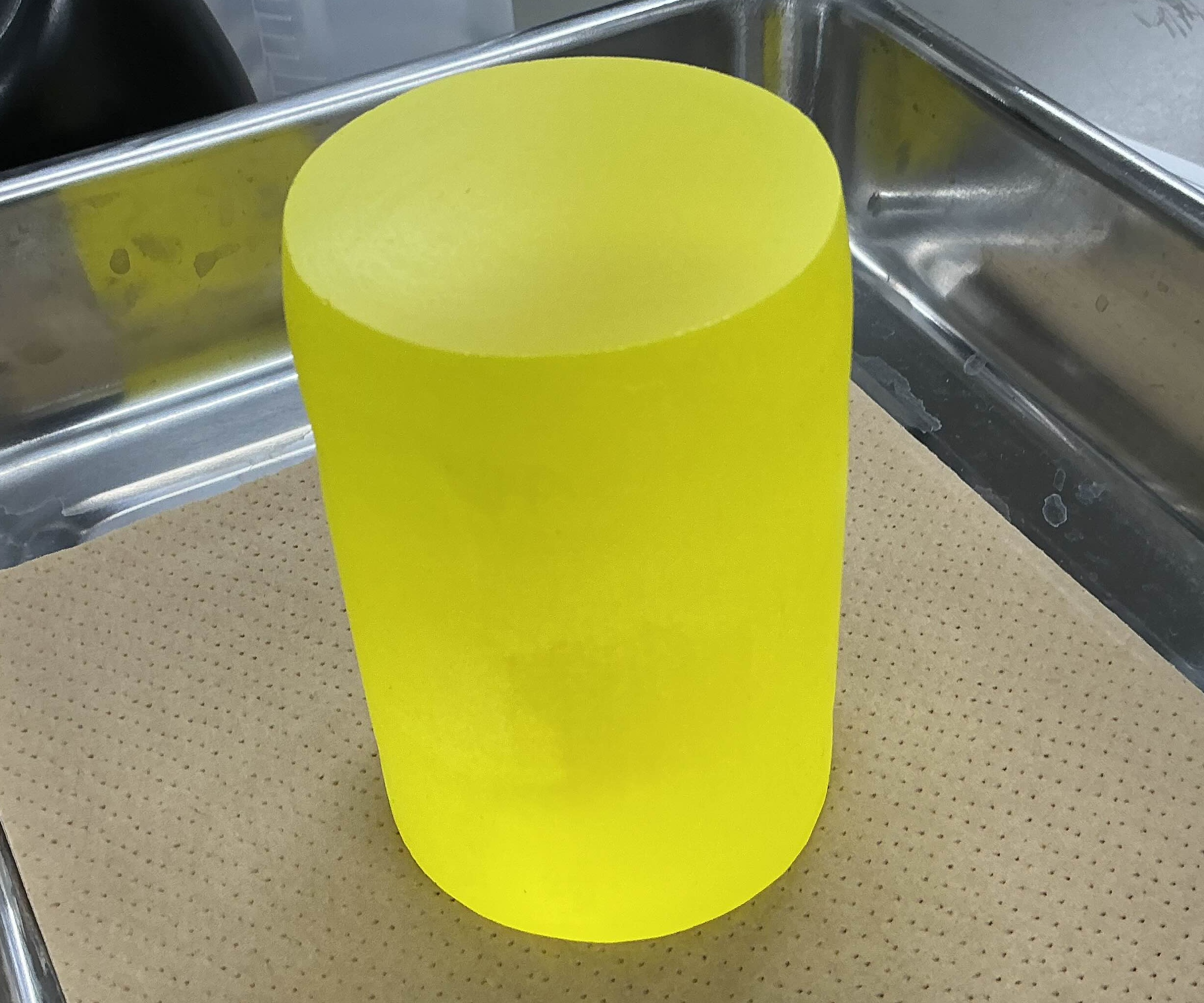}
  \caption{GAGG scintillator used in the PIKACHU experiment. The crystal has a diameter of 65~mm, a length of 100~mm, and a weight of 2.412~kg.}
  \label{fig:Crystal}
\end{figure}

Figures~\ref{fig:Experiment} shows a schematic diagram and a photograph of the CCT setup constructed in this study. The test detector is the GAGG scintillator shown in Fig.~\ref{fig:Crystal}, which has a diameter of 65~mm and a length of 100~mm, and was grown from the purified raw materials described in Ref.~\cite{Omori2024}. The scintillation light was read out by a photomultiplier tube (PMT, R6231-100, Hamamatsu Photonics~\cite{R6231}) through an acrylic light guide (LG). The LG has the shape of a truncated cone with diameters of 65~mm and 50~mm at its two ends, matching the coupling surfaces of the GAGG scintillator and the PMT, respectively. The interfaces between the LG, PMT, and GAGG scintillator were coupled using optical grease. The GAGG scintillator and LG were wrapped with PTFE tape as a reflector, and together with the PMT, were covered with black tape for light shielding. The bias voltage applied to the PMT was $-1200$~V. The detector configuration and measurement conditions are identical to those employed in the actual double beta decay search of the PIKACHU experiment.

In this study, a germanium (Ge) detector (AEGIS-GX40 manufactured by Mirion Technologies~\cite{AEGIS}) was employed as a reference detector because it exhibits superior energy resolution, 0.26\% ($\sigma$) at 662~keV, compared with that of the GAGG scintillator, 4.47\% ($\sigma$) at 662~keV, as well as a sufficient proportional response for $\gamma$ rays, with a residual non-linearity of less than 0.17~keV over the energy range of 276--2614~keV. The energy resolution and non-linearity quoted above were evaluated from the photopeaks of $\gamma$-ray sources prior to the CCT measurements. The Ge detector is equipped with a Ge crystal with a diameter of 61.7~mm and a length of 59.7~mm. A bias voltage of $-2509$~V was applied to the Ge detector. The signal was processed using trapezoidal shaping in the module and subsequently output to the data acquisition (DAQ) system described below.

The positions of these detectors were arranged to define $\theta$, and the distance between the detector surfaces was optimized within the range of 10--100~mm considering the lead blocks placed to prevent direct irradiation of the Ge detector from the $\gamma$-ray sources.

The GAGG and Ge signals were processed using a DAQ system constructed with Nuclear Instrument Module (NIM) electronics, as shown in Fig.~\ref{fig:Experiment_setup}. Both signals were split using a linear fan-in/fan-out (FIFO) module for direct input to a flash ADC (FADC) digitizer and for trigger generation. The FADC digitizer was a DT5720 manufactured by CAEN~\cite{CAEN}, which recorded waveforms with a sampling rate of 250~MS/s, a resolution of 12~bits, and a dynamic range of 2~V$_{\mathrm{pp}}$, where V$_{\mathrm{pp}}$ denotes the peak-to-peak input voltage. The GAGG signal used for triggering was first amplified by a factor of 10 using a photomultiplier amplifier (P.A.), and then further amplified by a factor of 5 with a shaping time of 150~ns using a shaping amplifier (S.A.). The amplified signal was discriminated (Disc.) with a threshold of $-30$~mV, generating a 5~$\mu$s gate using a gate generator (Gate). This GAGG signal was required to be in coincidence with the Ge signal, which was discriminated with a threshold of $+50$~mV, after which a 12~$\mu$s gate was opened. The coincidence signal was then used as the trigger for the FADC digitizer.

The sources and energies of the incident $\gamma$ rays, together with the scattering angles employed in the CCT measurements, are summarized in Table~\ref{tab:datasets}. The $^{22}$Na source was a standard check source with an activity of $8.9\times10^{5}$~Bq, calibrated one week before the CCT measurements. The $^{208}$Tl source used in this study originated from a radioactive isotope contained in a 2$\%$ thoriated tungsten electrode used for tungsten inert gas welding. The expected energies of the Compton electrons, $E_{\gamma}-E^{\prime}_{\gamma}$ which was calculated using Eq.~(\ref{eq:eq1}), are also listed in the table. The coincidence events were acquired at approximate rates of 1--2~Hz for the $^{22}$Na and 0.1~Hz for the $^{208}$Tl datasets.

\begin{table}[htbp]
  \centering
  \caption{Summary of the datasets obtained in the CCT measurements. The quantities $E_{\gamma}$ and $E^{\prime}_{\gamma}$ represent the energies of the incident and scattered $\gamma$ rays, respectively, and $\theta$ denotes the Compton scattering angle defined in Eq.~(\ref{eq:eq1}). The values of $E_{\gamma}-E^{\prime}_{\gamma}$ were calculated using Eq.~(\ref{eq:eq1}).}
  \label{tab:datasets}
  \begin{tabular}{ccccc}
    \hline
    \hline
    Source &
    $E_{\gamma}$ [keV] &
    $\theta$ [$^\circ$]&
    $E_{\gamma}-E'_{\gamma}$ [keV] &
    Measurement time [h]\\
    \hline

    \multirow{5}{*}{$^{22}$Na}
      & 511  & 30  & 60  & 6.0 \\
      & 1274 & 30  & 319  & 6.0 \\
      & 511  & 60  & 170  & 20.9 \\
      & 1274 & 60  & 707  & 20.9 \\
      & 1274 & 150 & 1049 & 21.4 \\
    \hline

    \multirow{2}{*}{$^{208}$Tl}
      & 2614 & 70  & 2016 & 0.7 \\
      & 2614 & 120 & 2313 & 28.1 \\
    \hline
    \hline
  \end{tabular}
\end{table}

In addition to the CCT measurements, the GAGG scintillator was irradiated with $\gamma$ rays from $^{22}$Na, $^{60}$Co, $^{133}$Ba, $^{137}$Cs, $^{208}$Tl, and $^{241}$Am to evaluate the nonproportional response for $\gamma$ rays. The photopeaks in the acquired spectra were subsequently analyzed.

\section{Analysis}
\label{sec:Analysis}

This section describes the procedures for event selection and energy calibration, followed by the evaluation of the nonproportional response.

\subsection{Event selection and energy calibration}
\label{sec:Event selection and energy calibration}
\begin{figure}[htbp]
  \centering
  \includegraphics[width=0.95\linewidth,height=0.28\textheight,keepaspectratio]{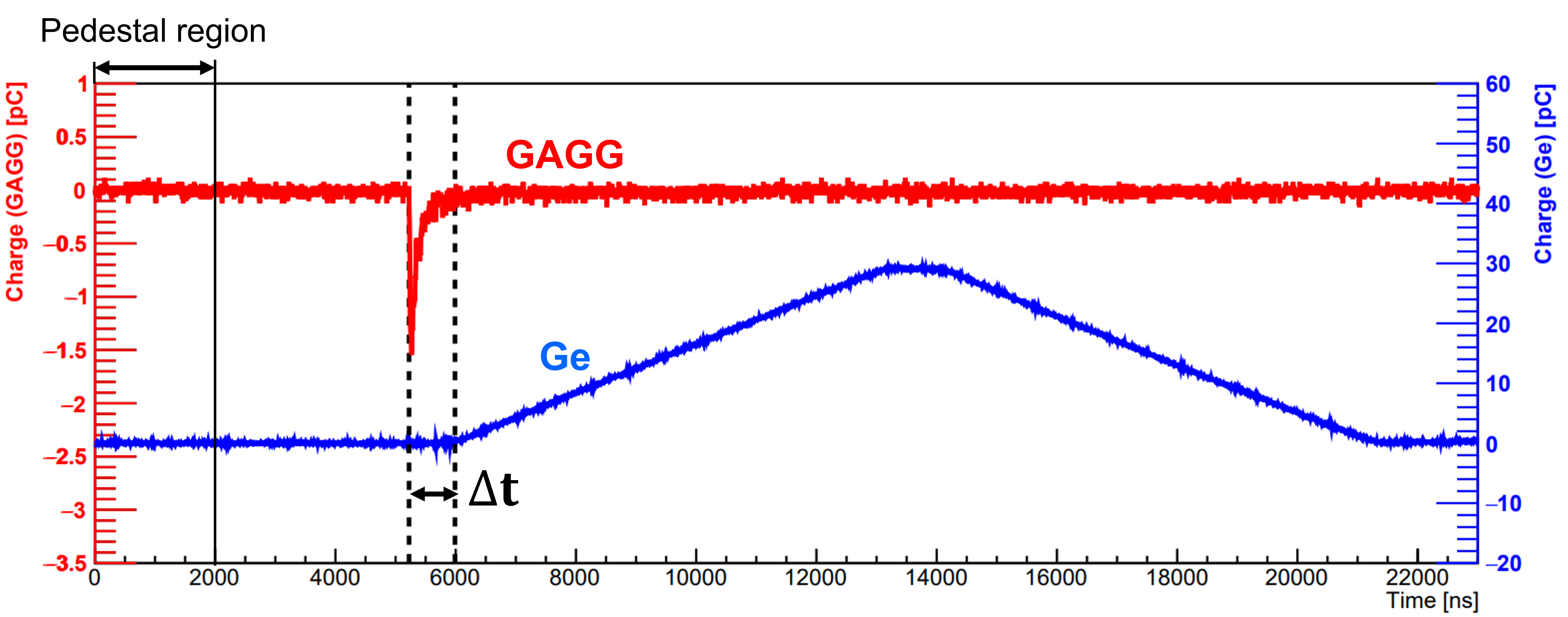}
  \caption{An example of the waveforms from the GAGG scintillator (red) and Ge detector (blue). The current values plotted on the vertical axis were calculated from the digitized waveform at each 4~ns sampling interval. The quantity $\Delta t$ is defined as the time interval from the rise time of the GAGG signal to that of the Ge signal, as indicated by black dashed lines. The region labeled ``Pedestal region'' indicates the interval used for the pedestal calculation.}
  \label{fig:waveform}
\end{figure}

Compton scattering events can be selected by requiring the time coincidence between GAGG and Ge signals. In this analysis, the time coincidence was determined from the time difference ($\Delta t$) defined as the time interval from the rise time of the GAGG signal to that of the Ge signal, as shown in Fig.~\ref{fig:waveform}. The rise time of the GAGG (Ge) signal was defined as the time at which the signal first exceeded the pedestal by $5\sigma$ for 10 (100) consecutive ADC samples. Each pedestal was calculated as the average ADC count over the first 2~$\mu$s of the pre-trigger region shown in Fig.~\ref{fig:waveform}. The events with $0\;\mathrm{ns}<\;\Delta t\;<2000\;\mathrm{ns}$ were selected as Compton scattering events.

The energies measured by the GAGG scintillator were determined from the waveform integral over a 1~$\mu$s window from the rise time and calibrated using zero-energy point (pedestal) and 1274~keV $\gamma$-ray peak in the $^{22}$Na spectrum. On the other hand, the energies measured by the Ge detector were determined from the waveform pulse height and calibrated using the $\gamma$-ray sources mentioned in the end of Sec.~\ref{sec:Experimental setup}.

\begin{figure}[htbp]
  \centering
  \subcaptionbox{$^{22}$Na, $\theta=30^\circ$\label{fig:2dhist_22na_30}}{%
    \includegraphics[width=0.32\linewidth]{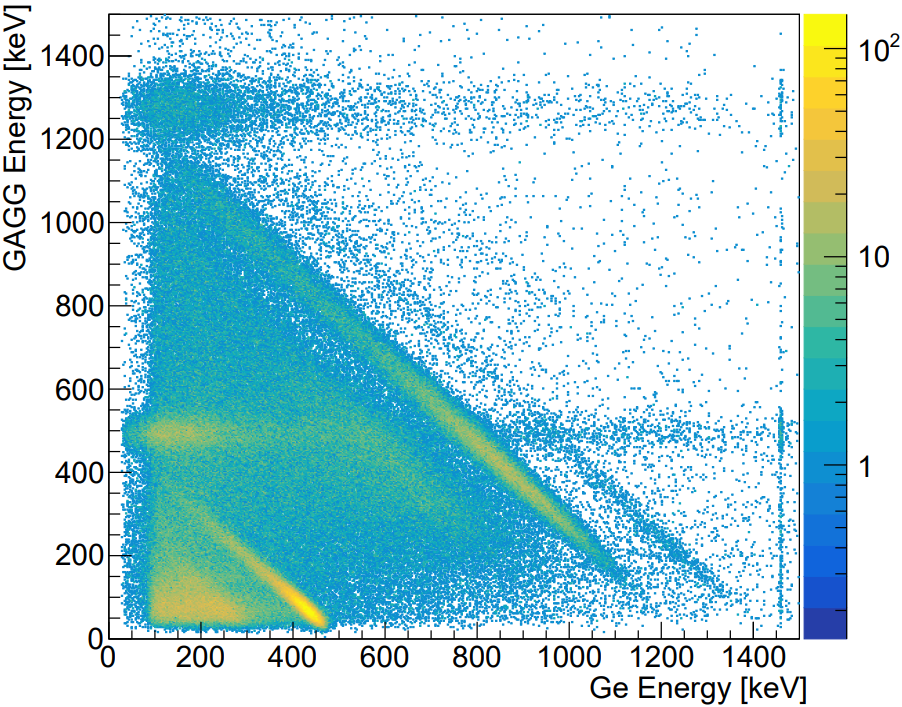}}
  \hfill
  \subcaptionbox{$^{22}$Na, $\theta=60^\circ$\label{fig:2dhist_22na_60}}{%
    \includegraphics[width=0.32\linewidth]{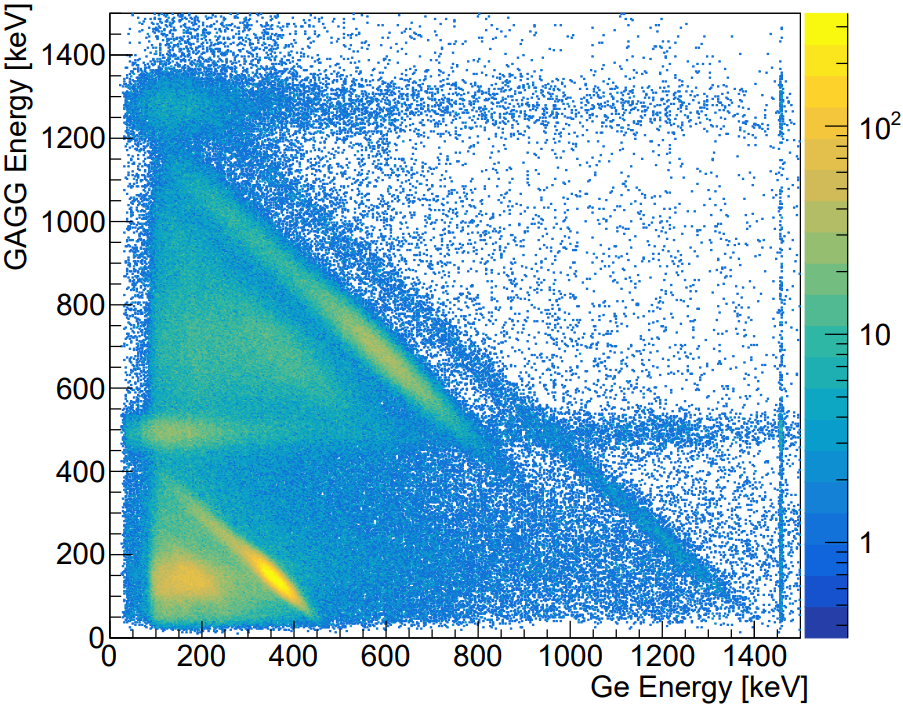}}
  \hfill
  \subcaptionbox{$^{22}$Na, $\theta=150^\circ$\label{fig:2dhist_22na_150}}{%
    \includegraphics[width=0.32\linewidth]{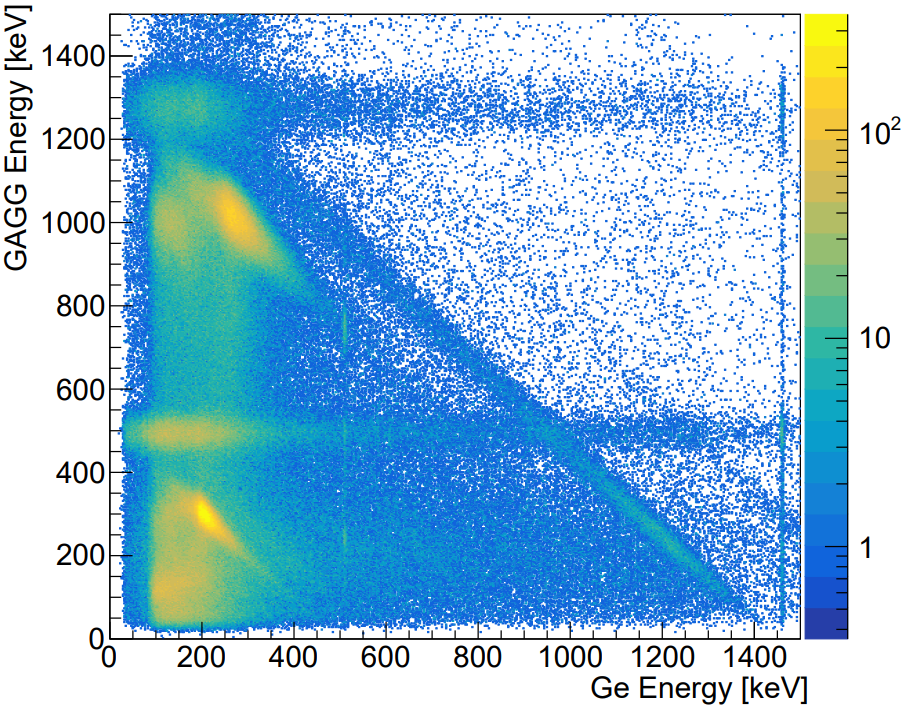}}
  \hfill
  
  \vspace{0.8em}

  \makebox[\linewidth][c]{%
    \subcaptionbox{$^{208}$Tl, $\theta=70^\circ$\label{fig:2dhist_208tl_70}}{%
      \includegraphics[width=0.32\linewidth]{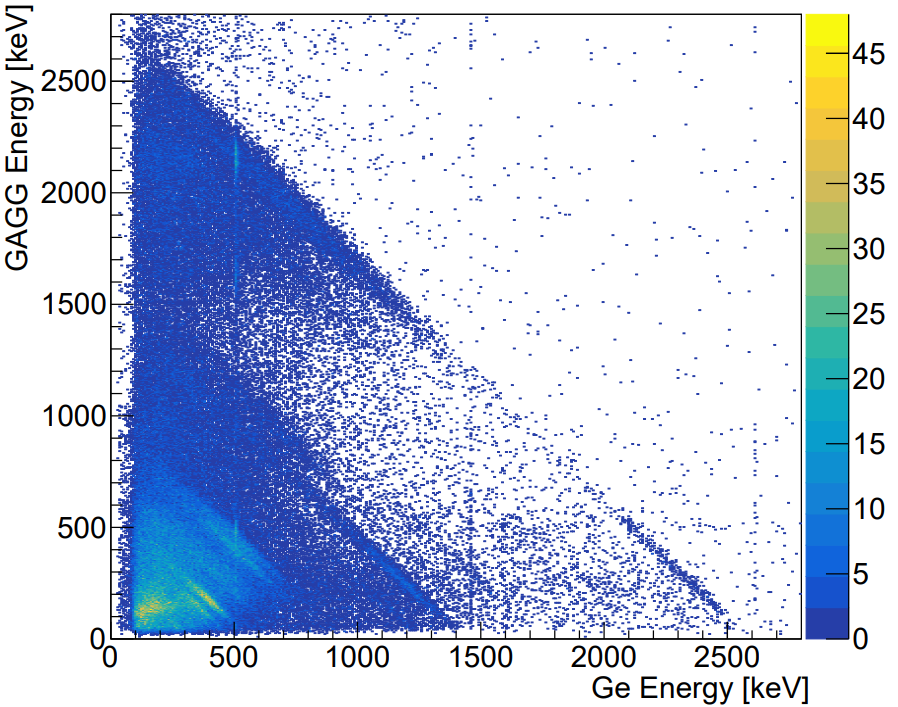}}%
    \subcaptionbox{$^{208}$Tl, $\theta=120^\circ$\label{fig:2dhist_208tl_120}}{%
      \includegraphics[width=0.32\linewidth]{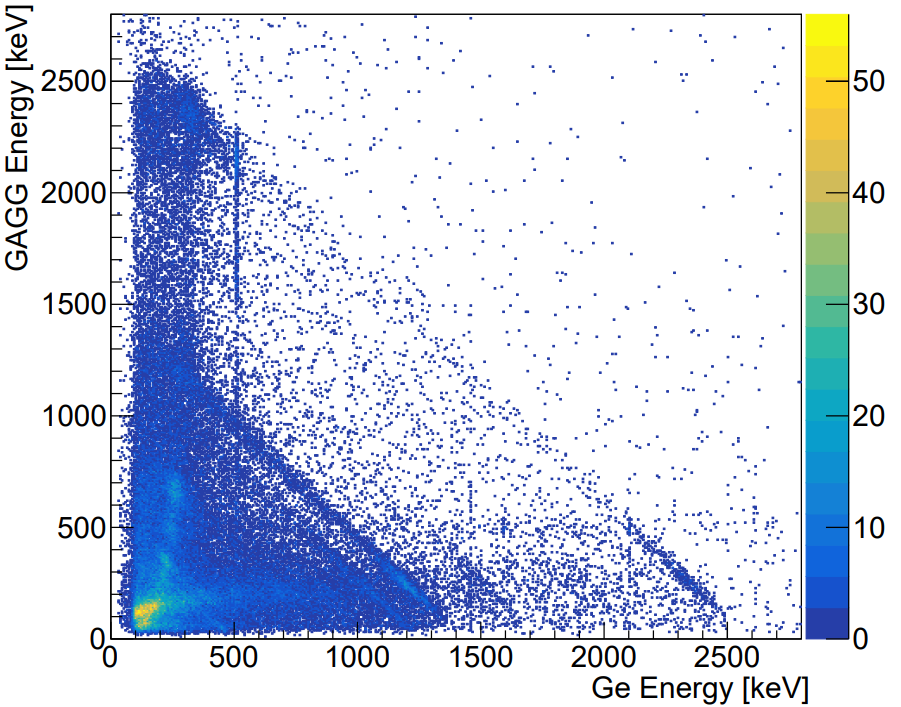}}%
  }
  \caption{Two-dimensional energy spectra for events selected using the $\Delta t$ condition described in the text. The horizontal and vertical axes represent the energies measured by the Ge detector and GAGG scintillator, respectively. The $z$-axes in panels (a)--(c) are shown on a logarithmic scale to make the Compton components easier to identify.}
  \label{fig:2Dhist}
\end{figure}

Figures~\ref{fig:2Dhist} show the two-dimensional energy spectra measured by the Ge detector ($x$-axis) and GAGG scintillator ($y$-axis) for each dataset after applying the $\Delta t$ coincidence selection described above. Events forming diagonal bands, for which the summed energies are approximately 511, 1274, or 2614~keV, are attributed to Compton scattering of the incident $\gamma$ rays. The diagonal bands are broader because the finite solid-angle acceptance of the experimental setup results in a wider distribution of the scattering angles of incident $\gamma$ rays. In contrast, events forming vertical or horizontal bands, as well as events distributed over other energy regions, correspond to accidental coincidences. Although Compton events induced by the 1460~keV $\gamma$ rays from environmental $^{40}$K were observed in all datasets, these events were excluded from the analysis.

\subsection{Evaluation of nonproportional response}
\label{sec:Evaluation of nPR}

In this study, the nonproportional response of the light yield of the GAGG scintillator was evaluated as the ratio of the measured energy to the deposited energy, which is denoted as nPR. In the CCT, the energy deposited in the GAGG scintillator can be calculated as the difference between the incident $\gamma$-ray energy and the energy of the scattered $\gamma$-ray measured by the Ge detector. Therefore, nPR can be expressed as

\begin{equation}
  \mathrm{nPR} = \frac{E_{\mathrm{GAGG}}}{E_\gamma - E_{\mathrm{Ge}}},
\label{eq:nPR}
\end{equation}

\noindent
where $E_{\mathrm{GAGG}}$ and $E_{\mathrm{Ge}}$ denote the energies measured by the GAGG scintillator and Ge detector, respectively. Thus, an nPR value of 1 indicates a proportional response normalized to the response for 1274~keV $\gamma$ rays, since the energy calibration of the GAGG scintillator was performed using the photopeak at 1274~keV as described in Sec.~\ref{sec:Event selection and energy calibration}.

To evaluate the energies measured by the GAGG scintillator for a given energy obtained from the CCT datasets, each two-dimensional spectrum shown in Fig.~\ref{fig:2Dhist} was divided into Ge-energy slices of 10, 20, and 50~keV for the 511, 1274, and 2614~keV Compton events, respectively. Although these slice widths are larger than the intrinsic energy resolution of the Ge detector, they were adopted to ensure sufficient statistics in each slice. By evaluating the nonproportional response using different slice widths, we confirmed that the obtained results showed no significant dependence on the slice width. For each slice, the corresponding GAGG energy spectrum was extracted, and the peak position obtained from the spectral fit was defined as $E_{\mathrm{GAGG}}$. On the other hand, the energy deposited in the GAGG scintillator, $E_{\gamma}-E_{\mathrm{Ge}}$, was determined from the corresponding $E_{\mathrm{Ge}}$, taken as the midpoint of each slice range in the Ge energy.

\begin{figure}[htbp]
  \centering
  \subcaptionbox{Small-nPR case (nPR=0.901): $^{22}$Na 511~keV ($\theta = 30^\circ$), Ge slice 430--440~keV.\label{fig:1Dhist_small}}{%
    \includegraphics[width=0.32\linewidth]{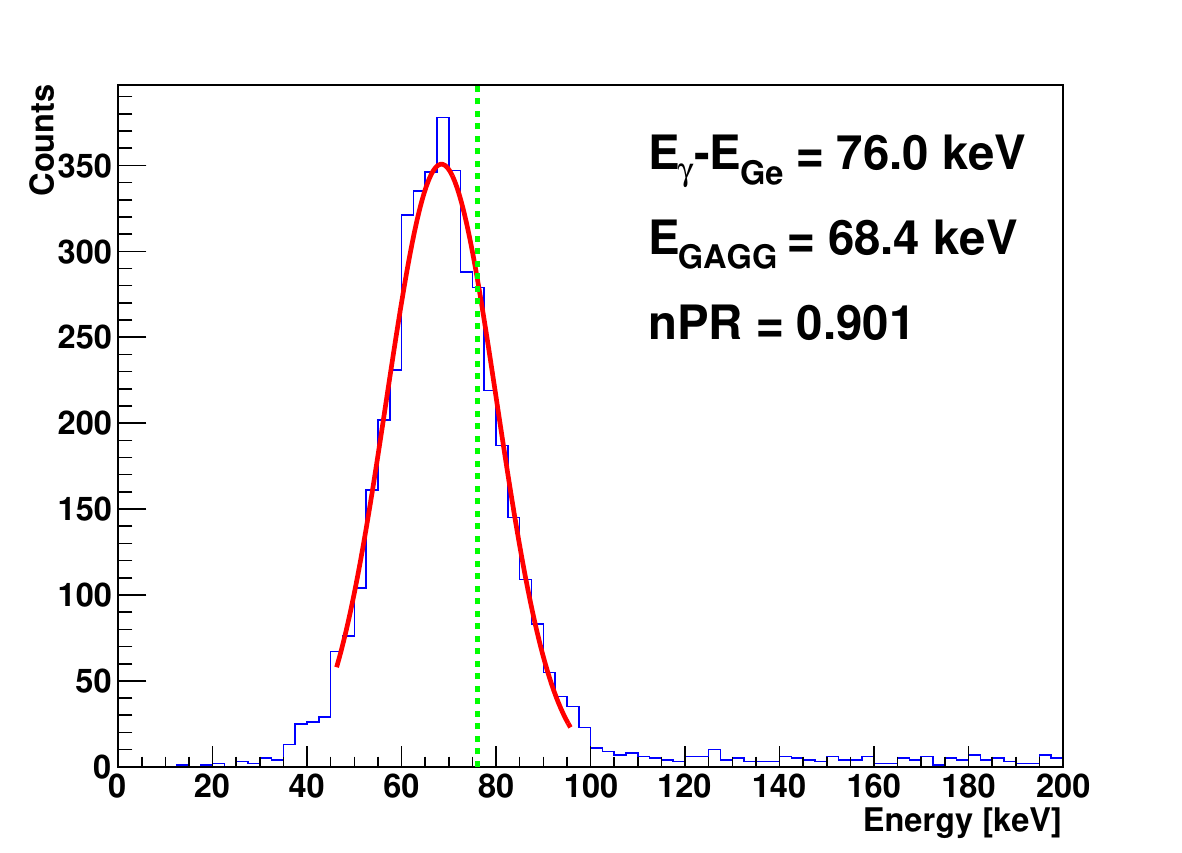}}
  \hfill
  \subcaptionbox{Proportional case: $^{22}$Na 1274~keV ($\theta = 60^\circ$), Ge slice 660--680~keV.\label{fig:1Dhist_equal1}}{%
    \includegraphics[width=0.32\linewidth]{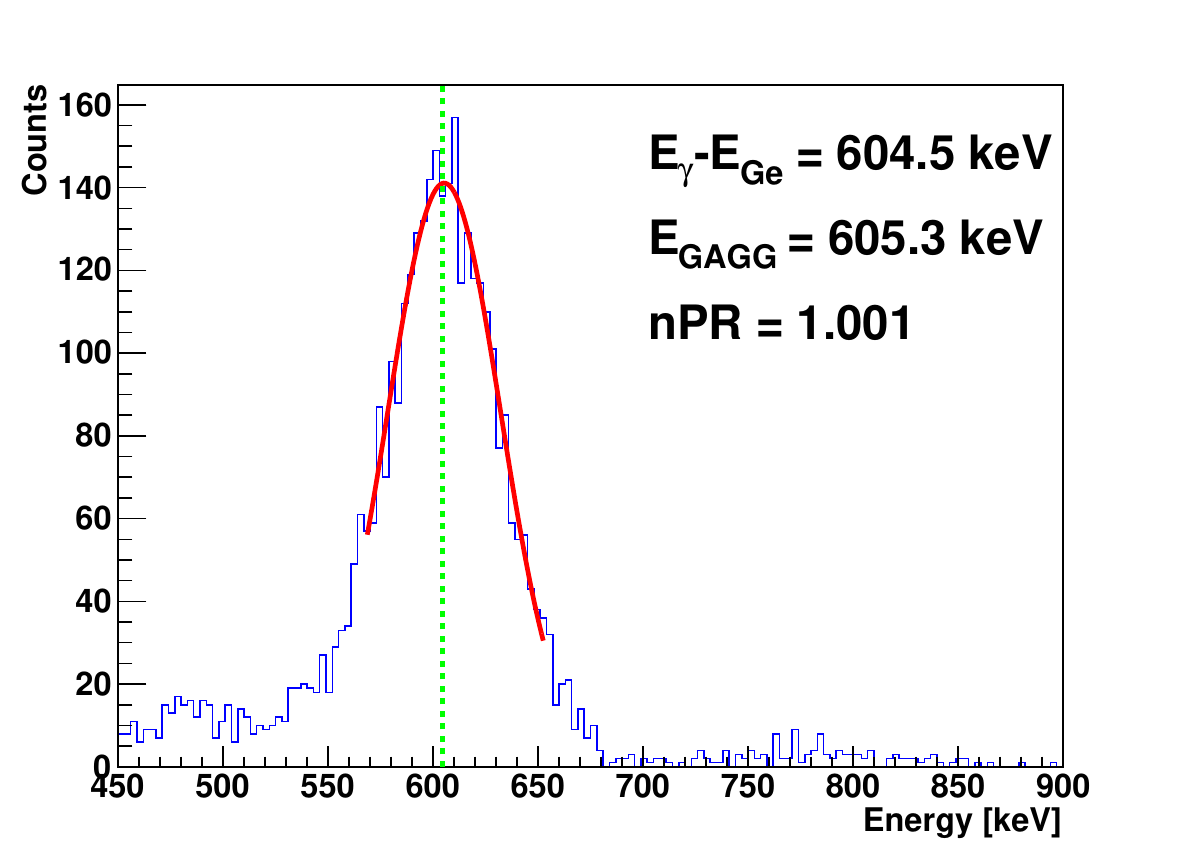}}
  \hfill
  \subcaptionbox{Large-nPR case: $^{208}$Tl 2614~keV ($\theta = 70^\circ$), Ge slice 800--850~keV.\label{fig:1Dhist_large}}{%
    \includegraphics[width=0.32\linewidth]{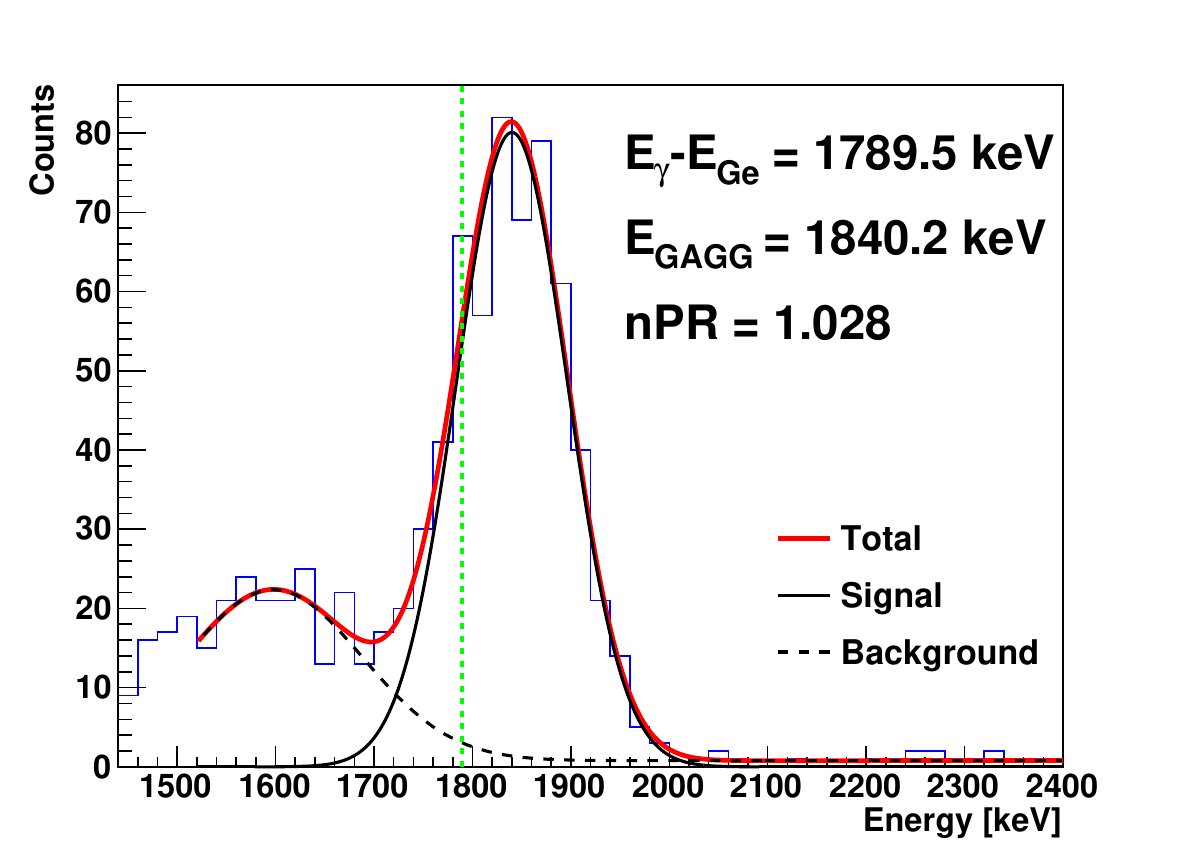}}
  \caption{Representative GAGG energy spectra for selected Ge-energy slices. The green dashed lines indicate the expected energy deposited in the GAGG scintillator. The red solid lines represent the fitting functions used to evaluate $E_{\mathrm{GAGG}}$. The black solid and dashed lines in panel (c) represent the signal and background components, respectively.}
  \label{fig:1Dhist}
\end{figure}

Figures~\ref{fig:1Dhist_small}, \ref{fig:1Dhist_equal1}, and \ref{fig:1Dhist_large} show GAGG energy spectra obtained from the 430--440~keV, 660--680~keV, and 800--850~keV Ge-energy slices for the $^{22}Na$ 511~keV ($\theta=30^\circ$), $^{22}Na$ 1274~keV ($\theta=60^\circ$), and $^{208}$Tl 2614~keV ($\theta=70^\circ$) datasets, respectively. The green dashed lines indicate the energies deposited in the GAGG scintillator, calculated from the corresponding $E_{\mathrm{Ge}}$. The spectra in Figs.~\ref{fig:1Dhist_small} and \ref{fig:1Dhist_equal1} were fitted with a Gaussian function (red lines), while the spectrum in Fig.~\ref{fig:1Dhist_large} was fitted with a double-Gaussian function (red line), consisting of a signal (solid black line) and a background (dashed black line) component. The background, which likely attributable to accidental coincidences, was empirically modeled using a Gaussian function because its distribution was approximately Gaussian within the fitting range. The resulting nPR values were confirmed to be insensitive to the inclusion of this Gaussian background component.

The same fitting procedure as shown in Figs.~\ref{fig:1Dhist} was applied to all datasets. Afterward, the nPR values were evaluated for each dataset around the deposited energy in the GAGG scintillator calculated from Eq.~(\ref{eq:eq1}), $E_{\gamma}-E^{\prime}_{\gamma}$, listed in Table~\ref{tab:datasets}. The resulting nPR values are presented in the next section.

\section{Results and discussion}

\begin{figure}[htbp]
  \centering
  \includegraphics[width=0.85\linewidth]{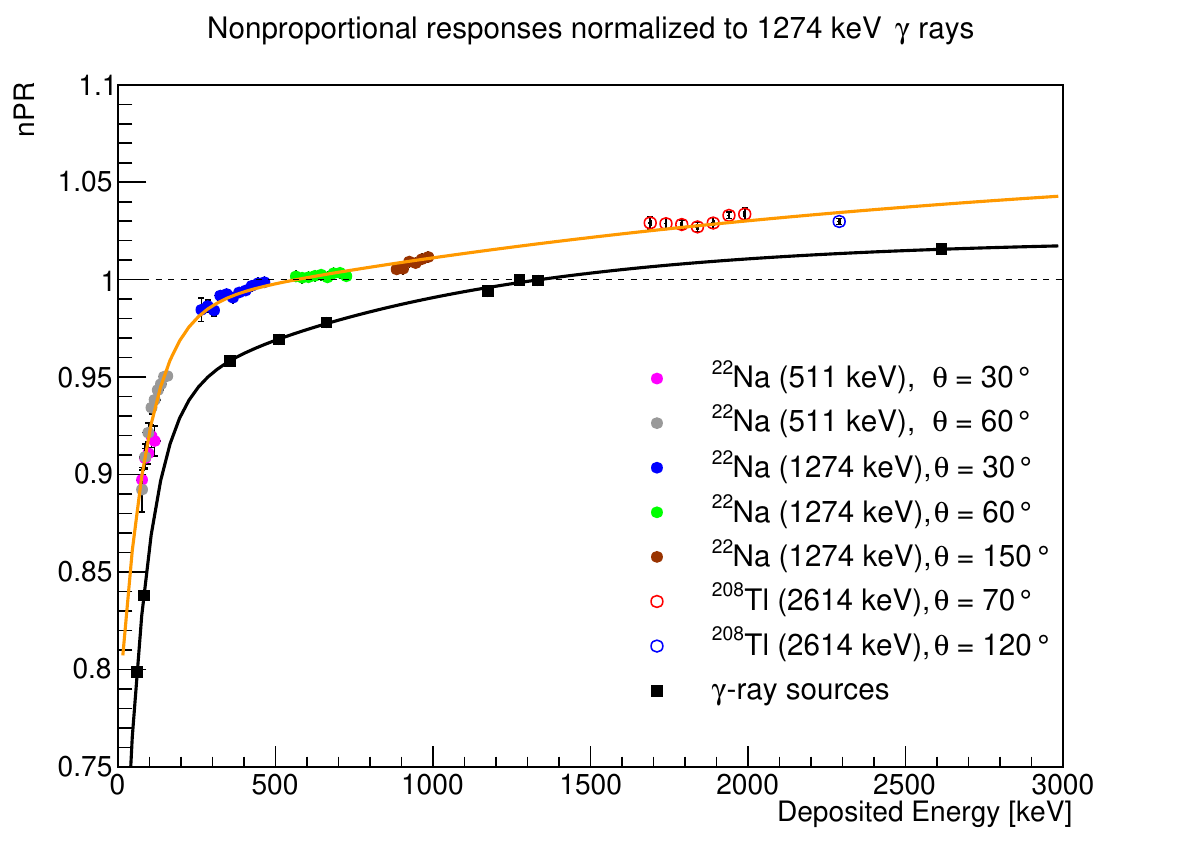}
  \caption{The nonproportional response of the GAGG scintillator. The nPR values for $\beta$ and $\gamma$ rays, normalized to the response for 1274~keV $\gamma$ rays, are shown as a function of the deposited energy by colored markers and black squares, respectively. The orange and black curves represent the corresponding fitted double-exponential functions.}
  \label{fig:Result1}
\end{figure}

The $E_{\mathrm{GAGG}}$ and $E_{\mathrm{Ge}}$ values were determined from the sliced spectrum analysis described in Sec.~\ref{sec:Evaluation of nPR}, and the nPR values for $\beta$ rays were subsequently calculated using Eq.~(\ref{eq:nPR}) for all datasets listed in Table~\ref{tab:datasets}. The colored markers in Fig.~\ref{fig:Result1} show the nPR values for $\beta$ rays as a function of the energy deposited in the GAGG scintillator. The filled and open markers represent the CCT data obtained using $^{22}$Na and $^{208}$Tl, respectively. The $x$-axis error bars correspond to the energy resolution of the Ge detector at each deposited energy, while the $y$-axis error bars reflect the uncertainty of the mean obtained from the fits of the GAGG spectra (e.g., Fig.~\ref{fig:1Dhist}). In addition, the nPR values for $\gamma$ rays, evaluated as a ratio of the measured photopeak energies to the corresponding incident $\gamma$-ray energies, are shown as black squares. Only the $y$-axis error bars are shown for these data, reflecting the uncertainty of the peak position obtained from the fit of the $\gamma$-ray spectra.

The orange and black curves represent double-exponential functions fitted to the nPR values for $\beta$ and $\gamma$ rays, respectively. The nPR values for $\beta$ and $\gamma$ rays asymptotically approach 1.07 and 1.02, respectively, at the deposited energies above 3000~keV. 

The nonproportional response obtained in this study below approximately 1000~keV is in good agreement with previously reported values for GAGG scintillators. For example, Ref.~\cite{Sibczynski2018} reports nonproportional response for $\gamma$ rays normalized to the response at 661.7~keV, whereas our results were originally normalized to 1274~keV. After renormalizing our data to the same reference energy, the nPR value at 100~keV for $\gamma$ rays was 0.88, consistent with Ref.~\cite{Sibczynski2018}. Likewise, the nPR value for $\beta$ rays at 100~keV, normalized to the response at 435.4~keV $\beta$ rays, was 0.93, in agreement with Ref.~\cite{Kaewkhao2016}. On the other hand, the nonproportional response of GAGG scintillators to radiations above 1000~keV has not been reported previously and is presented for the first time in this study. Figure~\ref{fig:Result1} shows a stronger nonproportional response for $\gamma$ rays than for $\beta$ rays in the GAGG scintillator. Such differences between the responses for $\beta$ and $\gamma$ rays have also been reported for Ce:LaBr$_3$ and Ce:LYSO ($\mathrm{(Lu,Y)_2SiO_5}$), whose nonproportional responses were measured using the CCT with a Ge detector~\cite{Swiderski2012}. Such a stronger nonproportional response for $\gamma$ rays is attributed to photoelectric absorption by inner-shell electrons. In this process, the absorbed energy is redistributed into multiple low-energy electrons through X-ray and Auger-electron cascades. As a result, a larger fraction of the energy is carried by low-energy electrons, which enhances the influence of low-energy nonproportional response compared with the case of Compton electrons, where the energy is primarily deposited by a single high-energy electron.

\begin{figure}[htbp]
  \centering
  \subcaptionbox{$\alpha$-ray background spectrum. \label{fig:Results2_alpha}}{%
    \includegraphics[width=0.495\linewidth]{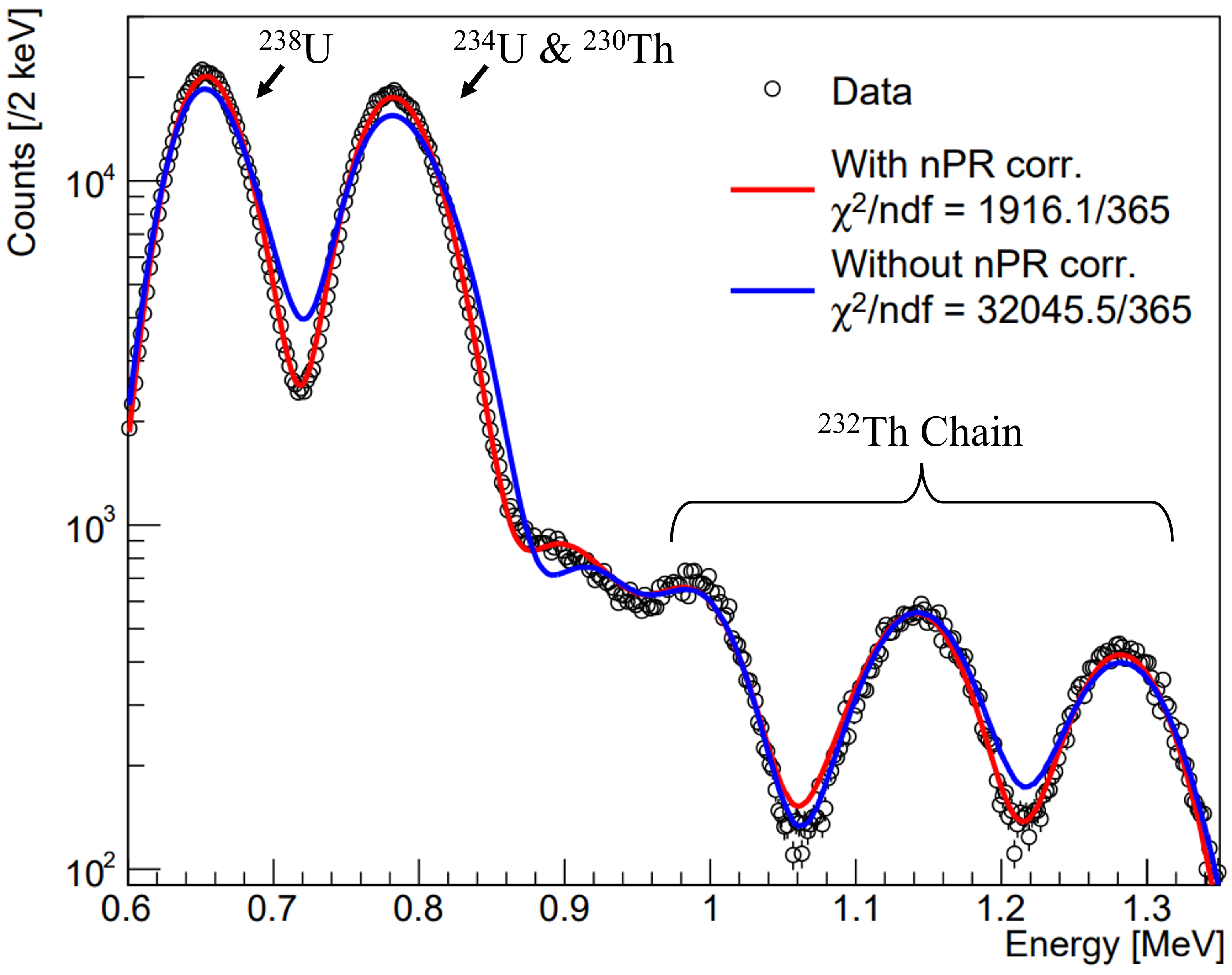}}
  \hfill
  \subcaptionbox{$\beta(\gamma)$-ray background spectrum.\label{fig:Results2_beta}}{%
    \includegraphics[width=0.495\linewidth]{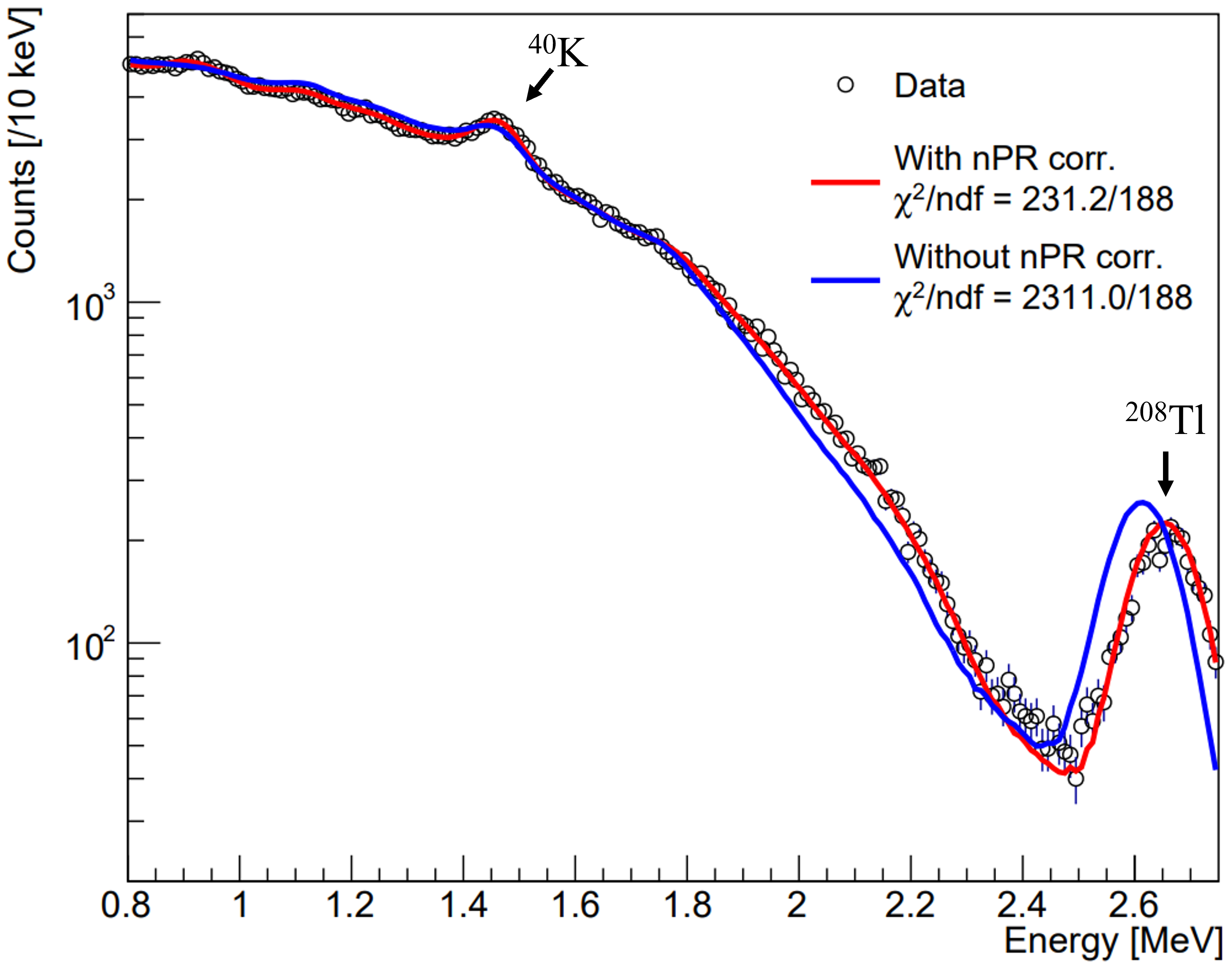}}
  \caption{Background-model fits to the measured $\alpha$- and $\beta$($\gamma$)-ray spectra. The black points represent the background spectrum of the GAGG scintillator measured at the Kamioka underground laboratory. The blue and red curves correspond to the models without and with the nonproportional response correction, respectively. Only the total models are shown; the individual background components are omitted for clarity. The radionuclides responsible for the major peaks are also indicated in the figure.}
  \label{fig:Results2}
\end{figure}

By using the fitting results obtained with double-exponential functions, the nonproportional responses for $\beta$ and $\gamma$ rays was incorporated separately into the background model for the PIKACHU experiment, which is a GEANT4-based background model incorporating the contributions of the uranium/thorium decay chains, $^{40}$K, $^{176}$Lu, and external gamma radiations. Such separate treatment is also adopted in liquid-scintillator experiments such as KamLAND-Zen (see Figure 6.13. in Ref.~\cite{Matsuda2016}). Figures~\ref{fig:Results2} show the fits to the $\alpha$- and $\beta$($\gamma$)-ray background spectra obtained with the background model. The black points represent the background spectrum measured with the GAGG scintillator shown in Fig.~\ref{fig:Crystal} over 264 hours of data taking at the Kamioka underground laboratory. The energy in Fig.~\ref{fig:Results2_alpha} is expressed in terms of the gamma-equivalent energy, which is attributable to the scintillator quenching effect for $\alpha$ particles. The blue and red curves represent the model without and with the nonproportional response measured in this study, respectively. The fitting ranges for $\alpha$- and $\beta$($\gamma$)-ray spectra is 0.60--1.35~MeV and 0.80--2.75~MeV, respectively. The chi-square values shown in Figs.~\ref{fig:Results2} indicate that the model with correction of the nonproprotional response provides a more accurate reproduction of the data.

For example, the $\alpha$-ray background includes de-excitation $\gamma$ rays emitted immediately after the $\alpha$ decay to an excited state of the daughter nucleus. Therefore, the nonproportional response for $\gamma$ rays affects the fitting quality of the $\alpha$-ray spectrum. On the other hand, for the $\beta$($\gamma$)-ray background, the nonproportional response for $\beta$ rays has the largest impact because $^{234\mathrm{m}}$Pa, which emits $\beta$ rays with a $Q$ value of 2.269~MeV, is the dominant background component in the GAGG scintillator~\cite{Omori2026}. Consequently, the discrepancy around 2.0~MeV observed in the model without the nonproportional response in Fig.~\ref{fig:Results2_beta} is significantly reduced in the updated model. In addition, the $\gamma$-ray peaks around 2.6~MeV originating from $^{208}$Tl outside the GAGG scintillator are also reproduced correctly.

\section{Conclusion}

In this study, the nonproportional response of the GAGG scintillator used in the PIKACHU experiment was evaluated for both $\beta$ and $\gamma$ rays. The response for $\beta$ rays was measured using the CCT employing a Ge detector, while that for $\gamma$ rays was evaluated using several monoenergetic $\gamma$-ray sources.

The measurement results reveal that the GAGG scintillator exhibits a strong nonproportional response for $\gamma$ rays than for $\beta$ rays within the measured energy range of 50--2614~keV. The nonproportional responses were incorporated into the background model of the PIKACHU experiment as double-exponential functions of the deposited energy. The model incorporating the nonproportional response successfully reproduced both $\alpha$- and $\beta$($\gamma$)-ray background data, compared with the model without the correction. These results demonstrate that the understanding of the nonproportional responses over a wide energy range provides important input for accurate energy calibration and detector response modeling in scintillator-based experiments.

The measurement technique established in this study can be directly applied to the evaluation of the nonproportional response of future GAGG scintillators developed for the PIKACHU experiment.

\section*{Acknowledgments}
This work was supported by JSPS KAKENHI Grant No. 25H02168, 23H01196 and 23K25892.


\begin{thebibliography}{9}
\bibitem{Danevich2001}
F. A. Danevich \textit{et al.},
\textit{Nucl. Phys. A} \textbf{694}, 375-391 (2001).
\href{https://doi.org/10.1016/S0375-9474(01)00983-6}{10.1016/S0375-9474(01)00983-6}.

\bibitem{Kochurikhin2020}
V. Kochurikhin \textit{et al.},
\textit{J. Cryst. Growth} \textbf{531}, 125384 (2020).
\href{https://doi.org/10.1016/j.jcrysgro.2019.125384}{10.1016/j.jcrysgro.2019.125384}.

\bibitem{Kamada2016}
K. Kamada \textit{et al.},
\textit{IEEE Trans. Nucl. Sci.} \textbf{63}, 443 (2016).
\href{https://doi.org/10.1109/TNS.2016.2521399}{10.1109/TNS.2016.2521399}.

\bibitem{Omori2024}
T. Omori \textit{et al.},
\textit{Prog. Theor. Exp. Phys.} \textbf{2024}, 033D01 (2024).
\href{https://doi.org/10.1093/ptep/ptae026}{10.1093/ptep/ptae026}.

\bibitem{Sibczynski2018}
P. Sibczy\`{n}ski \textit{et al.},
\textit{Nucl. Instrum. Methods Phys. Res. A} \textbf{898}, 24-29 (2018).
\href{https://doi.org/10.1016/j.nima.2018.03.050}{10.1016/j.nima.2018.03.050}.

\bibitem{Kaewkhao2016}
J. Kaewkhao \textit{et al.},
\textit{Appl. Radiat. Isot.} \textbf{115}, 221-226 (2016).
\href{https://doi.org/10.1016/j.apradiso.2016.06.030}{10.1016/j.apradiso.2016.06.030}.

\bibitem{Omori2026}
T. Omori \textit{et al.},
\textit{Nucl. Instrum. Methods Phys. Res. A} \textbf{1082}, 171023 (2026).
\href{https://doi.org/10.1016/j.nima.2025.171023}{10.1016/j.nima.2025.171023}.

\bibitem{R6231}
Hamamatsu Photonics K.K.
\textit{Photomultiplier tubes and assemblies for scintillation counting and high energy physics},
\href{https://www.hamamatsu.com/content/dam/hamamatsu-photonics/sites/documents/99_SALES_LIBRARY/etd/High_energy_PMT_TPMZ0003E.pdf}{https://www.hamamatsu.com/}.

\bibitem{AEGIS}
MIRION TECHNOLOGIES
\textit{Aegis Portable HPGe Spectrometer Data Sheet},
\href{https://assets-mirion.mirion.com/prod-20220822/cms4_mirion/files/pdf/spec-sheets/aegis-data-sheet.pdf?_gl=1*13jux4r*_gcl_au*NTk0MjYyNTUxLjE3Nzc2MTE4ODA.*_ga*NDU1MzYwOTMzLjE3Nzc2MTE4ODA.*_ga_GMYBLJ5Q7G*czE3Nzg2NTg3NDEkbzUkZzEkdDE3Nzg2NTg3NzAkajMxJGwwJGgxNzMwMzg5NTk1}{https://www.mirion.com/}.

\bibitem{CAEN}
CAEN 
\textit{CAEN DT5720 Product page},
\href{https://www.caen.it/products/dt5720/}{https://www.caen.it/products/dt5720/}.

\bibitem{Swiderski2012}
L. Swiderski \textit{et al.},
\textit{IEEE Trans. Nucl. Sci.} \textbf{59}, 303 (2012).
\href{https://doi.org/10.1109/TNS.2011.2175407}{10.1109/TNS.2011.2175407}.

\bibitem{Matsuda2016}
S. Matsuda,
``Search for Neutrinoless Double-Beta Decay in $^{136}$Xe after Intensive Background Reduction with KamLAND-Zen,'' PhD thesis, Tohoku University (2016).
\href{https://www.awa.tohoku.ac.jp/Thesis/ThesisFile/matsuda_sayuri_d.pdf}{https://www.awa.tohoku.ac.jp/}.

\end{thebibliography}
\end{document}